\documentclass[sigconf, nonacm]{acmart}
\AtBeginDocument{%
}

\usepackage{graphicx}
\usepackage{textcomp}
\usepackage{bmpsize}
\usepackage{xcolor}
\usepackage{lipsum}
\usepackage{hyperref}
\usepackage{listings}

\usepackage{algorithm}
\usepackage{algpseudocode}
\usepackage{amsmath}
\usepackage{graphicx}

\usepackage{subcaption}

\PassOptionsToPackage{breaklinks=true,unicode=true,urlcolor=blue,colorlinks=true,citecolor=red,linkcolor=red}{hyperref}
\usepackage{hyperref}

\usepackage{multirow}
\usepackage{array}
\newcolumntype{P}[1]{>{\centering\arraybackslash}p{#1}} 

\usepackage{enumitem}

\usepackage{multicol}
\usepackage{multirow}

\usepackage{tabularx}
\usepackage{balance}
\usepackage{epstopdf}

\usepackage{amsfonts}

\usepackage{tcolorbox} 

\usepackage[nameinlink,capitalize]{cleveref} 

\usepackage[font=small]{caption}

\newcommand{\paragraphb}[1]{\vspace{0.75ex}\noindent{\bf #1.} }

\newcommand{\ignore}[1]{}

\definecolor{orange}{RGB}{255,127,80}
\definecolor{darkgreen}{RGB}{50,127,0}
\definecolor{Blue}{RGB}{0,0,255}

\usepackage{comment}
\usepackage{amsmath}

\usepackage{changepage}

\newtheorem{takeawaythm}{Takeaway}

\usepackage{comment}

\newcounter{mynote}[section]
\newcommand{\notecolor}{blue}
\newcommand{\thenote}{\thesection.\arabic{mynote}}

\newcommand{\wbnote}[1]{\ifx\outforreview\undefined\refstepcounter{mynote}{\it\textcolor{orange}{(WB~\thenote: { #1})}}\fi}
\newcommand{\mjnote}[1]{\ifx\outforreview\undefined\refstepcounter{mynote}{\it\textcolor{darkgreen}{(MJ~\thenote: { #1})}}\fi}
\newcommand{\vbnote}[1]{\ifx\outforreview\undefined\refstepcounter{mynote}{\it\textcolor{\notecolor}{(VB~\thenote: { #1})}}\fi}

\newcommand{\fixme}[1]{\ifx\outforreview\undefined\textbf{\textcolor{red}{[FIXME: #1]}}\fi}
\newcommand{\todo}[1]{\ifx\outforreview\undefined\textbf{\textcolor{red}{[TODO: #1]}}\fi}

\usepackage[T1]{fontenc}

\usepackage{graphicx}
\usepackage{booktabs} 
\usepackage{multirow}
\usepackage{listings}
\begin{document}

\title{Thwart Me If You Can: An Empirical Analysis of Android Platform Armoring Against Stalkerware}

\author{Malvika Jadhav}
\affiliation{%
  \institution{University of Florida}\country{}}
\email{jadhav.m@ufl.edu}
\author{Wenxuan Bao}
\affiliation{%
  \institution{University of Florida}\country{}}
\email{wenxuanbao@ufl.edu}
\author{Vincent Bindschaedler}
\affiliation{%
  \institution{University of Florida}\country{}}
\email{vbindschaedler@ufl.edu}

\begin{abstract}
    Stalkerware is a serious threat to individuals' privacy that is receiving increased attention from the security and privacy research communities. Existing works have largely focused on studying leading stalkerware apps, dual-purpose apps, monetization of stalkerware, or the experience of survivors. However, there remains a need to understand potential defenses beyond the detection-and-removal approach, which may not necessarily be effective in the context of stalkerware.

In this paper, we perform a systematic analysis of a large corpus of recent Android stalkerware apps. We combine multiple analysis techniques to quantify stalkerware behaviors and capabilities and how these evolved over time. Our primary goal is understanding: how (and whether) recent Android platform changes --- largely designed to improve user privacy --- have thwarted stalkerware functionality; how stalkerware may have adapted as a result; and what we may conclude about potential defenses. Our investigation reveals new insights into tactics used by stalkerware and may inspire alternative defense strategies.



\end{abstract}


\maketitle

\section{Introduction}\label{sec:intro}
Stalkerware is an unfortunate example of technology enabling intimate partner violence and stalking. Stalkers or intimate partner abusers may install a {\em stalkerware} app on a victim's mobile device, allowing them to remotely monitor activity without their knowledge or consent. This threat now affects individuals across the globe, as the usage of stalkerware has grown explosively in recent years~\cite{kaspersky2023stalkerware}. 

There is substantial prior research on various aspects of stalkerware. For example, prior research~\cite{parsons2019predator,chatterjee2018spyware} has highlighted the data collection features of stalkerware, including the collection of text messages, web searches, and real-time location. 
There have also been efforts in characterizing the detectability of stalkerware~\cite{roundy2020many,han2021towards}, its monetization strategies~\cite{gibson2022analyzing}, dual-use applications~\cite{almansoori2022global,chatterjee2018spyware,almansoori2024web,stephenson2023s}, and other considerations~\cite{stephenson2023s}. Additional work has explored forensic tools, detection frameworks, and covert access behaviors \cite{MANGEARD2024301677,bonam2025current,hewitt2025unveiling}.

Despite our growing understanding of stalkerware that stems from this prior research, there remains unanswered questions regarding defense mechanisms. Looking at stalkerware as a type of malware, suggests that we merely need to detect it and remove it from devices where it is found. But looking at stalkerware considering the unique feature of the adversary model reveals a more complicated picture. 

Stalkerware apps may be installed on the victim's device surreptitiously or without their consent by an intimate partner abuser with physical access to the device. For example, abusers have been known to gift devices with stalkerware pre-installed to their victims ~\cite{securelistStateStalkerware}. Furthermore, even if the victim becomes aware that a stalkerware app is installed on their device, they may be unable or unwilling to remove it because that could alert the abuser. Studies such as Spitzburg and Cupach~\cite{spitzberg2003mad} show that stalkers' behavior may escalate if their methods are unsuccessful. Therefore, removing stalkerware could put victims at greater harms.

In this paper, we seek to improve our understanding of mitigation against stalkerware. Inspired by efforts to harden the Android platform model in recent years, we investigate whether these changes are effective in mitigating stalkerware. Google has previously targeted stalkerware by enacting new terms of service for the Google Play Store in October 2020 that explicitly prohibit such software. Our study was conducted prior to the introduction of platform changes in Android 14 and beyond, such as stricter controls on accessibility services for sideloaded apps, that are expected to have a significant impact on stalkerware. As such, our evaluation reflects the state of mitigation mechanisms up to and including Android 13.

Our investigation leverages a corpus of 8,428 Android stalkerware app samples collected until December 2022 by the Coalition Against Stalkerware. This is a larger corpus than that used in prior studies, and our analysis using VirusTotal's first seen date reveals that 6646 (78.99\%) appeared after 2020 suggesting that the corpus captures the latest stalkerware trends. Moreover, because this corpus is longitudinal in nature, we can correlate capabilities and behaviors with specific changes in the Android permission system to determine what impact (if any) a particular change had on the stalkerware landscape. The corpus also contains several stalkerware app families --- apps that share the same package ID but have different app samples, presumably released at different times --- which allows us to study how stalkerware capabilities evolved within a given app family.

The findings of our investigation are mixed. While we found clear and widespread evidence of stalkerware reacting to recent changes in Android, we found little evidence that this reduced their capabilities or hindered their functionality. For example, Android 11 introduced the READ\_PHONE\_NUMBERS permission to replace the broader READ\_PHONE\_STATE permission. This change aimed to limit the scope of access to phone numbers and enhance user privacy. Despite this change, 136 out of 335 app samples that target Android 11 and above within our corpus still perform call logging. 

The problem is multi-faceted, and we find that stalkerware apps have numerous ways to evade restrictions from the Android platform that would impede their functionality. Many such restrictions can be avoided by granting the appropriate permissions, changing app settings during the time of installation, or by enabling the apps to have device policy controller privileges like making them Device Administrator apps. 

In fact, we found in our corpus a small subset of apps containing no malicious behaviors (and requesting no permissions) except for instructions guiding stalkers through the installation process of stalkerware apps on victims' devices. These installation guides provide detailed instructions on how to side-load stalkerware apps, disable settings and notification to keep stalkerware undetected, and grant the required permissions to ensure stalkerware apps' functionality is unimpeded. 

Android’s recent changes will likely force stalkerware developers to update their apps and provide increasingly detailed instructions to enable installation. Nevertheless, new defense approaches may be needed to continue addressing stalkerware threats.

\paragraphb{Structure of the paper}
\cref{sec:related} provides background on Android and stalkerware, and discusses related work. \cref{sec:methodology} describes our corpus and details our analysis methodology. \cref{sec:corpus} provides an overview of the corpus and discuss overall trends in the evolution of stalkerware capabilities. \cref{sec:changes} discusses changes of the Android platform model that affect stalkerware, and \cref{sec:resilience} analyzes tactics by which stalkerware avoid having their functionality restricted. \cref{sec:discussion} reflects on our findings and discuss ideas for alternative mitigation strategies. We discuss limitations and ethical considerations for our study in~\cref{sec:limits}. Finally, we conclude and highlight directions for future research in~\cref{sec:conclusions}.

\section{Background and Related Work}\label{sec:related}
\subsection{Background}

\paragraphb{Android}
It is an open-source operating system by Google for mobile devices like smartphones and tablets and supports app development in Java, Kotlin, and C++. Its critical permission system regulates app access to device resources and sensitive user data, such as location, contacts, and media, ensuring user privacy and security. Developers of Android apps are required to declare the necessary permissions for their application in the manifest file ({\tt AndroidManifest.xml}) that is part of the app package~\cite{androidapiref}. Upon installation or during run time, users are prompted to grant or deny these permissions. 

Android apps are distributed and installed using an Android application package (APK), compressed files that bundle necessary resources, and are mainly available through the Google Play Store. Users can also sideload these apps via APKs from third-party app stores. A typical APK file is composed of the following:
\begin{itemize}[leftmargin=1.5em,nolistsep,noitemsep]
    \item {\tt AndroidManifest.xml}: the manifest file contains indispensable information, such as package name, version, required permissions, and declared components (activities, services, content providers, and broadcast receivers).
    
    \item {\tt Classes.dex}: this contains the app's Java or Kotlin code in DEX (Dalvik Executable) format. 

    \item {\tt resources.arsc}: this bundle compiled app resources such as strings, colors, and styles.

    \item {\tt res}: This directory contains various resource files, including images, XML layouts, and raw assets, organized into subdirectories based on their types and configurations (e.g., drawable, layout, and values).

    \item {\tt lib}: The lib directory contains native libraries, typically written in C or C++, that the app relies on for specific functionality or performance optimization.
    
\end{itemize}

The methods used by researchers to analyze Android apps fall into two classes: static analysis \cite{li2017static,payet2012static,li2016reflection} and dynamic \cite{alzaylaee2017improving,yerima2019machine} analysis. These methods can help researchers reverse engineer an app to understand its behavior, identify its potential security vulnerabilities, or investigate privacy implications.

Static analysis involves examining the app's source code, resources, and configuration files without executing it. For Android apps, a common step is decompilation~\cite{mauthe2021large, fora2014beginners, nolan2012decompiling, desnos2011android} which is used to convert the compiled DEX bytecode into human-readable Java or Kotlin source code. Tools such as JADX, apktool, and dex2jar are frequently utilized for static analysis. 
In contrast, dynamic analysis involves running the app in a controlled environment, such as an emulator, and monitoring its runtime behavior. This approach can help identify behaviors that may not be apparent through static analysis alone. %

For tasks such as taint analysis, researchers use tools such as FlowDroid~\cite{arzt2014flowdroid} and TaintDroid~\cite{enck2014taintdroid}, which are renowned for their efficacy in this domain. In our study, we also employed FlowDroid to conduct a comprehensive taint analysis on a subset of our dataset.

%
\paragraphb{Stalkerware}
Stalkerware refers to tools --- software programs, apps, and devices --- that enable someone to secretly spy on another person’s private life via their mobile device ~\cite{coalition1}.
Stalkerware technology facilitates intimate partner violence by allowing the abuser to monitor victim's sensitive information including location, text messages, photos, voice calls, and much more. In recent years, the problem of stalkerware has been on the rise worldwide. According to Kaspersky's 2023 report, stalkerware impacted users across 175 countries, with over 31,000 unique individuals affected worldwide~\cite{kaspersky2023stalkerware}. Often perpetrators may install such apps on a user's device with or without their consent and/or knowledge. In cases of stalkerware-enabled intimate partner violence (IPV), a key difference from other tech-enabled abuse is the abuser's access to the victim’s device. In some cases, the abuser gifts a device to the victim or their children, with stalkerware pre-installed without the victim's knowledge. Even if an IPV victim knows about the app, they are often powerless, as the abuser can confiscate the phone to reinstall it or coerce the victim into doing so. Although most stalkerware apps are specifically designed and marketed for people looking to spy on a cheating spouse or track their partner, abusers sometimes misuse legitimate apps, such as those meant for tracking lost devices or caregiving apps for the elderly, for stalking. This is often due to the easy availability of such legitimate apps in the app stores. App stores have drafted strict policies to ban such apps, but a large portion of stalkerware apps are also distributed directly through the app's website.

\paragraphb{What is targetSDKVersion and why it matters}
Google Play uses rules to filter what apps are visible to a user browsing or searching for applications from the Google Play app. Usually, the filters are mentioned in the Android manifest file. One such filter is the \texttt{uses-sdk} tag used to specify the \texttt{minSDKVersion} and \texttt{targetSDKVersion} of an Android app. \texttt{minSDKVersion} specifies the lowest SDK version compatible with an app. To ensure that an app functions correctly with a specific \texttt{targetSDKVersion}, it needs to be tested against that version before declaration. Even when the API level of a device is higher than the version declared by an app in its manifest file, the system enables compatibility behaviors so that the app continues to work as expected \cite{leveldoc}. The \texttt{targetSDKVersion} field is more important for apps published on the Google Play store and may not have an effect on apps distributed outside of Google Play. Even if new Android API levels are introduced, changes are typically additive and if any methods are to be discontinued, they are usually deprecated and not removed so that older apps can still keep functioning as intended. 


\subsection{Related Work}

\paragraphb{Spyware and Stalkerware}
There is a plethora of recent work on tracking, spyware, and stalkerware. Razaghpanah et al.~\cite{razaghpanah2018apps} study the mobile tracking ecosystem. They proposed an automated way to detect third-party ads and track traffic. They also discuss business practices that involve data sharing with subsidiaries and third-party affiliates. In a different vein, Štefanko et al.~\cite{vstefanko2021android} identified vulnerabilities in 86 Android stalkerware apps, including server issues, application problems, and network leakage. Recently, Gibson et al.~\cite{gibson2022analyzing} studied the monetization strategies of Android stalkerware developers. They find a wide range of payment services, including Google's own in-app billing, subscription models, and cryptocurrencies. Other recent work provides detailed analyses of some stalkerware apps. For example, Heasley et al.~\cite{Androidstalkerware} maintain a repository that details the functionality of various Android stalkerware apps. Roundy et al.~\cite{roundy2020many} proposed CreepRank, a novel algorithm to identify and characterize creepware, or apps used for interpersonal attacks. Their analysis of a large dataset of mobile apps uncovered various forms of creepware with harmful intentions. Liu et al. \cite{liu2023no} analyzed consumer Android spyware apps, focusing on their functionality and security vulnerabilities. They studied 14 leading apps, highlighting their monitoring methods and evasion techniques. The research points out the apps' privacy issues, especially insecure data transmission and storage, raising concerns about potential misuse. Rawat et al.~\cite{rawat2024exploring} analyzes the security risk in Android devices and identifies stalkerware as a significant security threat. It emphasizes proactive measures such as regular audits of app permissions and the use of anti-stalkerware tools to protect user privacy.

\paragraphb{Dual-use Apps}

One of the features that make studying stalkerware different than other malware is the extent of dual-use or dual-purpose apps. Dual-use apps refer to apps that have legitimate and potentially malicious uses. The primary usage of such kinds of application may be beneficial, but their function can also be used for malicious purposes. For example, ``find my phone'' -type apps can be used to find lost phones, but could also be used for intimate partner surveillance.

There is recent research focusing on dual-use apps or devices. For example, Chatterjee et al.~\cite{chatterjee2018spyware} studies the intimate partner surveillance (IPS) spyware ecosystem. They highlight the role of dual-purpose apps and explain why anti-virus and anti-spyware tools are insufficient. Almansoori et al.~\cite{almansoori2022global} conducted a survey on the prevalence of dual-use Android applications that are employed for IPS across 15 languages and 27 countries. They identified 854 dual-use apps after collecting over 51,000 apps from the Google Play store. ~\cite{maier2025surveillance} further reveal that many sideloaded parental-control apps function as covert surveillance tools, often transmitting sensitive data without encryption and exhibiting stalkerware-like behavior.
Almansoori et al. \cite{almansoori2024web} investigate the availability and efficacy of online resources for survivors of IPS. The study found that while abusers can easily access detailed guides and tools for conducting IPS, survivors often encounter online resources that are poor, inaccurate, and lack actionable advice. This disparity highlights a significant gap in support available to survivors compared to the resources that assist abusers.
Stephenson et al.~\cite{stephenson2023s} focus on dual-use IoT devices, identifying 39 instances of IoT-Enabled Intimate Partner Abuse. As remedies, they suggest enhancing transparency for IoT devices, revising IoT access control protocols, and heightening awareness of IoT abuse.

\paragraphb{Particularly related works}
Particularly related works include Almansoori et al.~\cite{almansoori2022global}, Parsons et al.~\cite{parsons2019predator}, Han et al.~\cite{han2021towards}, and Mannan et al. \cite{mannanprivacy}. Almansoori et al.~\cite{almansoori2022global} only focus on dual-use apps, whereas we study stalkerware more broadly. Parsons et al.~\cite{parsons2019predator} focuses on the consumer spyware ecosystem, including stalkerware apps, and the marketing strategies of stalkerware developers. They also discussed the high-level capabilities of stalkerware and provided a taxonomy. Our focus in this paper is different as we aim to provide quantitative evidence of fine-grained capabilities and behavior of Android stalkerware apps using a large corpus and to understand how changes to Android have affected stalkerware functionality. Han et al.~\cite{han2021towards} conduct a technical analysis of stalkerware apps, focusing on capability detection and identification. Their study uses a dataset of 1462 apps, while our analysis covers a larger set and additionally considers platform-level changes over time. Mannan et al.~\cite{mannanprivacy} investigate the privacy and security implications of stalkerware in the context of intimate partner violence. Their study highlights vulnerabilities, evaluates detection tools, and examines the supporting third-party ecosystem. Our work complements these efforts by examining a broader set of 8421 apps while emphasizing the evolving impact of Android platform policies on stalkerware functionality.

\section{Methodology and Data}\label{sec:methodology}
\subsection{Dataset}\label{sec:data}
We use a dataset of Android stalkerware samples obtained from the Coalition Against Stalkerware. The Coalition Against Stalkerware is an international alliance of partners including IT security companies, domestic violence survivor networks, and digital rights advocacy groups~\cite{coalition}. The coalition maintains a Stalkerware Threat List (STL) consisting of malware samples identified as stalkerware. 

\subsection{Analysis Process}
\label{sec:analysis}
We decompile samples using the JADX decompiler and JEB Pro to obtain their (decompiled) source code. We perform our analysis in two phases: static and dynamic. We focus on static analysis in the first phase to provide comprehensive coverage over our entire corpus. We then perform taint analysis using Flowdroid~\cite{arzt2014flowdroid} on a subset of our corpus to juxtapose the results obtained from our query-based approach with the outcomes from Flowdroid. We complement this analysis with dynamic analysis to validate the findings of the static analysis. A secondary motivation of dynamic analysis is to study the instructions given to the adversaries and to recognize distinct unique apps communicating with the same backend servers.

\paragraphb{Static Analysis}
While most stalkerware apps exhibit similar traits, their range of capabilities varies. To structure our analysis, we start from the taxonomy developed by Parsons et al.~\cite{parsons2019predator}, but tailor it to our specific requirements. We mainly focus on nine capabilities, namely location tracking, (SMS and MMS) messages monitoring, call logs, contacts, phone call recordings, calendar event tracking, keylogging, data exfiltration, and social media tracking. Based on this taxonomy, we compiled a comprehensive collection of APIs, methods, and permissions linked to each capability. \cref{tbl:APIcap} lists the permissions required for each capability in our taxonomy.

We perform static analysis in two phases:

For the \textbf{first phase} of static analysis of the corpus, we focus on permission usage and API call patterns. We develop a set of Python scripts to query the decompiled source code, manifest files, and other APK resource files. Specifically, we develop regex-based queries and utilize code tree parsing libraries, similar to XML parsers, to facilitate tasks such as permissions enumeration, API usage tracking, and network library identification. To ensure that our analysis was sound and complete, we crafted precise regular expressions, adhering closely to the official Android documentation, to identify specific permissions, code patterns related to systems API calls, and other strings indicative of stalkerware functionality. To minimize the risk of undercounting or overcounting, we employ a twofold cross-validation approach:
\begin{enumerate}
\item Perform dynamic analysis of all unique apps in our corpus.
\item Compare our method's results with Taint analysis using Flowdroid on a random sample of apps.
\end{enumerate}

For permissions, we identify the tags: \texttt{<permission>} and 

\texttt{<uses-permission>} within the manifest file of each app using an XML parser. It is worth noting that research on Android apps has shown~\cite{felt2011android} it is not uncommon for app developers to request permissions they do not use, so we may expect this to also hold true of stalkerware.

To understand the capabilities of stalkerware related to permissions and data exfiltration, we focus our API analysis process on identifying function calls to built-in Android methods. This approach is crucial because apps can introduce redundant user-defined functions that ultimately remain unused, so relying solely on the presence of user-defined methods is potentially misleading. Therefore, we identified specific Android APIs, as detailed in~\cref{tbl:APIcap}, which are linked to the nine capabilities that we focus on. Accessibility services are aimed at helping disabled users, as they can read what is happening on the screen and perform actions like clicking or entering text for the user. Hence, we have quantified such attempts to monitor the changes in screen content within our corpus to identify the apps that are using accessibility service to monitor social media apps. For the social media monitoring capability we have considered both the scenarios: Apps that require a rooted device to directly access social media data bases such as msgstore.db (WhatsApp) and apps that use accessibility services to monitor screen content related to these apps. We then manually trace the actual execution flow of these APIs for a random sample of 20 unique apps, ensuring that we only considered calls that were intended for data exfiltration at the end.

In the \textbf{second phase} of static analysis we focus on analysis of user visible strings that are mainly present in the resources section of every decompiled APK. User visible strings refer to the strings displayed on the User Interface of an app to the device user, for example warnings, information about requested permissions, and more. We begin our analysis by parsing the xml file for each app sample to isolate the textual data relevant to a set of target keywords and phrases. Specifically keywords associated with instructions about Android versions, battery optimization or rooting the device, as well as app samples' capabilities. To reduce the risk of undercounting or overcounting, we employ the following steps:
\begin{enumerate}
\item Use exact matching and fuzzy matching (e.g. Levenshtein distance, regular expressions) to recognize keywords with variations in spelling or formatting.
\item Perform dynamic analysis of all unique apps in our corpus installing each app on an emulator.
\end{enumerate}

\begin{table}[t!]
\centering
\caption{Android versions of samples in our corpus. Note that developers may deliberately target older SDK versions.}\label{tbl:targetversions}
{
 \begin{tabular}{ccc}
 \toprule
 Platform Version &  Target SDK Version  &  Samples (\%)\\ \midrule
Android 9 & 28   & 2390 (78.61\%) \\ 
Marshmallow & 23 & 1358 (16.12\%) \\ 
Lollipop & 22  & 524 (6.22\%) \\ 
Oreo & 26  & 464 (5.51\%) \\ 
Lollipop & 21 & 448 (5.32\%) \\ 
Android 10 & 29 & 315 (10.36\%) \\ 
Nougat & 25 & 309 (3.67\%) \\  
Android 11 & 30 & 291 (9.57\%) \\ 
Oreo & 27 & 133 (1.57\%) \\  
Nougat & 24  & 46 (0.55\%) \\ 
Android 12, 13 & 31-33  & 44 (1.45\%)\\ 
Snow Cone (Android 12L) & 32 & 14 (\%) \\ 
Tiramisu (Android 13) & 33 & 8 (\%) \\ 
Android 4 and below & 4 to 20 & 533 (6.32\%)\\
\bottomrule
\end{tabular}
}
\end{table}

\begin{table*}[]
\begin{tabular}{ccccc}
\hline
\multirow{2}{*}{\begin{tabular}[c]{@{}c@{}}Number of  API \\ levels targeted ($n$)\end{tabular}} & \multirow{2}{*}{Unique Apps} & \multicolumn{2}{c}{  Commonly possessed capabilities}                                  & \multirow{2}{*}{\begin{tabular}[c]{@{}c@{}}Targeted API level \\ (Highest frequency)\end{tabular}} \\ \cline{3-4}
                                                                                           &                              & Top 1    & Top 5                                                 &                                                                                                    \\ \hline
2                                                                                          & 66                           & Location & Location, SMS, Call Recording, Call Logging, Facebook & 28 (36.9\%)                                                                                        \\ \hline
3                                                                                          & 9                            & Location & Location, SMS, Call Recording, Call Logging, Contacts & Below 20 (44\%)                                                                                    \\ \hline
4                                                                                          & 5                            & Location & Location, SMS, Call Recording, Call Logging, Facebook & 29 (22.9\%)                                                                                        \\ \hline
5                                                                                          & 4                            & Location & Location, SMS, Call Recording, Facebook, SMS          & 29 (45.6\%)                                                                                        \\ \hline
6                                                                                          & 1                            & Location & Location, SMS, Call Recording, Call Logging, Facebook & 23 (47.5\%)                                                                                        \\ \hline
7                                                                                          & 2                            & Location & NA                                                    & 30 (25\%)                                                                                          \\ \hline
\end{tabular}
\caption{Targeted API levels and capabilities. Number of  API levels targeted refers to the count of distinct API levels targeted by a unique app (considering all different versions of the same app in the corpus). Same app means app samples with the same unique identifier. The unique apps column is the count of unique apps in the corpus that target $n$ distinct API levels.}
\label{tbl:evolution}
\end{table*}

\section{A First Look at The Corpus}
\label{sec:corpus}

\subsection{Overview}
The Stalkerware Threat List (STL) obtained from the Coalition Against Stalkerware (CAS) we use for our study contains 8428 Android APK samples collected up to December 2022. Out of these app samples, 7 did not decompile properly, so we consider the remaining 8421 samples in our analysis.

We investigate the age of samples in our corpus using: (1) VirusTotal first seen time and (2) \texttt{targetSDKVersion.}

\paragraphb{VirusTotal}
We perform a systematic analysis of all APKs from our dataset using VirusTotal which is an online service that analyzes files and URLs for viruses, worms, trojans, and other types of malicious content.\footnote{\url{https://www.virustotal.com/gui/home/upload}}  We show the first seen time in \cref{tbl:virustotal_time}. The first seen time means the first data that someone uploads apps with the same hash to VirusTotal. It reveals that 50\% of these APKs first appeared post-2022. 

\begin{table}[th]
\centering
\caption{First seen time for the analyzed APKs and Detection rate for different antivirus engines over time. } 
\label{tbl:virustotal_time}
\resizebox{0.8\columnwidth}{!}{%
\begin{tabular}{ccc}
\toprule
Time & Number of APKs & Detection rate \\ \midrule
Unrecognized  & 720 (8.56\%) & NA  \\
Before 2020  & 1048 (12.46\%) & 23.60\% \\
2020-2022  & 2432 (28.90\%) & 30.21\% \\
After 2022  & 4214 (50.08\%) & 30.55\%   \\ \bottomrule
\end{tabular}%
}
\end{table}
In addition, we explore the detection capabilities of antivirus engines on VirusTotal in identifying malicious APKs. Our findings show that almost every APK successfully analyzed was flagged as malicious (there are 339 apps not recognized by VirusTotal). However, there is a notable variance in the detection rates among different antivirus engines and time. On average, only 29.38\% of these engines, out of an average total of 61, identified the sample app APK as malicious. This detection rate varied widely, with the lowest being just 1\% (except for not discovered), the highest reaching 85.33\%, and an average standard deviation of 12.87\%. These point to significant discrepancies in the detection efficacy of different antivirus engines. This highlights the importance of using multiple approaches for stalkerware detection and the ongoing need for comprehensive evaluation when evaluating potentially malicious apps.

\paragraphb{targetSDKVersion}
Successive versions of the Android platform (like Nougat, Pie, etc.) introduce updates in the framework APIs. However, API levels uniquely represent every such revision to the APIs~\cite{apilevels} and therefore we categorize samples accordingly. The distribution of API levels within our corpus is described in~\cref{tbl:targetversions}. This table specifically accounts for the \texttt{targetSDKVersions} for samples in our corpus. Here, it is important to highlight that the source code of 1566 samples lacks information about \texttt{targetSDKVersion}. Most of the samples in our corpus target Android 9 (API level 28 --- released in 2018). However, note that this does not mean that such apps were developed and published around 2018, since developers can opt to target SDK versions older than the most recent version at the time of development. In fact, VirusTotal detection results (\cref{tbl:virustotal_time}) post-2022 suggest either a significant lag in detection by VirusTotal engines in recognizing stalkerware or that stalkerware developers deliberately target older SDK versions.

\paragraphb{App Repackaging}
Since Android apps are uniquely identified by their package names specified in the manifest file~\cite{appid}, we can look for samples that are different versions of the same app. Within our corpus (subset) of 8421 app samples, we find 3143 distinct unique package names. Unless otherwise stated, we refer to unique samples as apps or {\em app samples} and unique package names as {\em unique apps}.

Aside from the natural evolution of apps where new samples with the same package name are created as developers update their app's functionality, we found evidence of app repackaging. App repackaging, in which the source code of an app is modified and redistributed~\cite{MERLO2021102181}. In our corpus, we came across a set of app samples that had the same application identifier but targeted different \texttt{targetSDKVersions}. Although we did not see any significant differences in capabilities, the instructions given to the adversary \footnote{In this paper, we use the term ``adversary'' to refer to the individual who installs or controls stalkerware on the target device (i.e., the perpetrator of surveillance).} within these apps were different when targeting different versions of Android. For example, a set of 21 apps target seven different API levels from 23 to 33 and all have the same application identifier. But we see that app samples from this set that target API levels 28 and above have an increasing set of instructions about how to disable battery optimization and grant permissions.

\begin{figure*}[ht]
    \centering
    \includegraphics[width=0.905\linewidth]{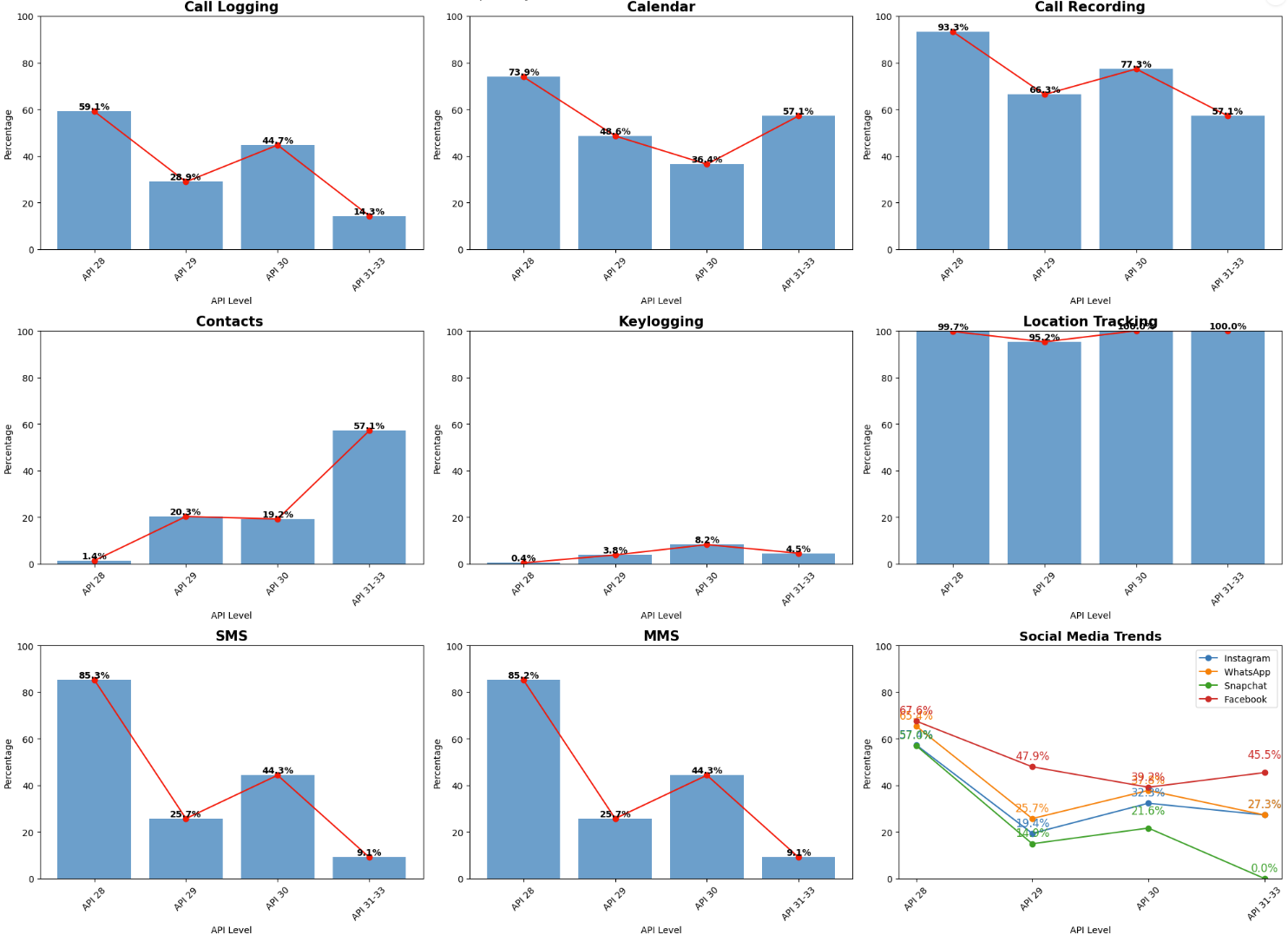}
    \caption{Stalkerware Capabilities across API levels: This figure demonstrates the trends in stalkerware capabilities across Android apps targeting different API levels (targetSDKVersion). For each capability, the barplot represents the percentage of apps that possess the capability within the set of apps targeting specified API level.}
    \label{fig:caps2}
\end{figure*}
\subsection{Stalkerware: Linking Motivation and Function}
In this section, we explore the motivations behind the use of Stalkerware and how these motivations are directly linked to the technical capabilities of these apps. \cref{tbl:APIcap} shows the number of samples and unique apps (unique application identifier) in the corpus that request permissions and the respective calls and libraries specific to each capability in our taxonomy. Throughout our discussions, when we say an app possesses a capability, it means the app requests the relevant permissions and utilizes the corresponding APIs (if at least one method of access is used) and methods in its codebase. This determination is based on the results of the static analysis using the method described in Section \ref{sec:analysis}

\paragraphb{Coercive Control} Location tracking is a pervasive capability in Stalkerware because it allows adversaries to restrict and control the victim. This is supported by the fact that 90.66\% (7635) of the apps in our corpus perform location tracking. 94\% of the apps that monitor location use the LocationManager API\footnote{\url{https://developer.android.com/reference/android/location/LocationManager}}, which can be used to track location even if a device does not have Google Play Services\footnote{Google Play services provides a system application called Google Play Protect Service that checks users' apps and devices for harmful behavior.
This means that an app can potentially avoid being flagged if Google Play Protect is absent and also facilitates distribution of the app for devices that do not come with pre-installed Google services}.  We find that 12.04\% (1014) apps in the corpus also monitor when users enter or exit predefined geographical zones using the Geofencing API. This capability is significant in our context, as it raises concerns about potential abuse. For example, an adversary might restrict a victim's access to help by setting up geofences around trusted contacts or support shelters. In addition, they could closely monitor if a victim leaves defined geofences, such as their home or office. These are just two examples, highlighting the broader potential for misuse.

\paragraphb{Persistent Access} Although not an obvious stalkerware functionality, we found that 600 (7.12\%) app samples in our corpus perform keylogging. Of these 600 app samples, 530 record user input text and, 144 track user notifications. The keylogging in stalkerware could be used to ensure persistent access to the user's device, even in scenarios where users change the device and other application passwords. Or, it could be used to steal user credentials to track the victim off-device.

\paragraphb{Social Isolation Tactics} By monitoring the interaction of a victim on social media and messaging platforms, an adversary can control the victim's social relationships and isolate them from their support network. We discovered two methods in which these stalkerware apps seem to capture user's social media data: (1) accessibility service\footnote{Accessibility Services are aimed at aiding users with disabilities, as they can read what is happening on the screen and perform actions like clicking or entering text for the user.}; and (2) default database of the social media app. 600 (7.12\%) apps in our dataset use accessibility services to capture content on the screen. We found that social media apps like Facebook, Instagram, Twitter, Musical.ly (now Tiktok), Pinterest, Zoom, and even Tinder, as well as browsers such as Chrome, Firefox, DuckDuckGo, and others, are monitored for user interactions like clicks, text changes, notifications, and scrolling. A small portion of app samples from our corpus also use the default databases of targeted social networks or messenger applications to capture a victim's social interactions. We manually examined a subset of 20 apps that query the default databases of apps like WhatsApp, Snapchat, Instagram, Facebook, Facebook Messenger, Kik, Line, Viber, and Telegram.  These apps first check whether a user's device is rooted since these databases are encrypted, and require root access for apps to bypass security measures. If the device is rooted, the apps then interact directly with the database using shell commands. The efficacy of tracking victim's message data through stalkerware apps may therefore hinge on whether the device is rooted or if the adversary can successfully root the device during the installation of the stalkerware app. \cref{tbl:apps-socialmedia-db} shows the count of app samples communicating with various social media and messenger app databases. From the table, we see that most apps monitoring messenger apps like WhatsApp, Line, Viber, and Telegram target Android 9 and above, which aligns with the rising popularity of these instant messengers for personal and sensitive communications. Along with capturing various social media and messenger applications, we found that stalkerware apps also capture SMS and MMS data. About 5932 (70.44\%) apps from our corpus possess the capability to capture SMS information. This is not surprising as we found a large number of app samples use SMS to receive commands as elaborated in~\cref{sec:commands}. 2955 (35.09\%) apps from the corpus capture MMS data such as sender and recipient particulars, images, videos, and audio messages. Furthermore, to capture even more details of the victim's day, we found 3523 (41.83\%) app samples also monitor calendar event details, including names, descriptions, locations, and timings.

\begin{table*}[th]
\begin{tabular}{llll}
\hline
\multirow{2}{*}{Application} & \multirow{2}{*}{Database}                                 & \multicolumn{2}{l}{Number of app samples} \\ \cline{3-4} 
                             &                                                           & Total           & API Level 28 +          \\ \hline
\multirow{2}{*}{Whatsapp}    & /data/com.whatsapp/databases/msgstore.db                  & 825             & 229                     \\ \cline{2-4} 
                             & /data/com.whatsapp/databases/wa.db                        & 295             & 235                     \\ \hline
Snapchat                     & /data/com.snapchat.android/databases/tcspahn.db           & 444             & 0                       \\ \hline
Instagram                    & /data/com.instagram.android/databases/direct.db           & 219             & 204                     \\ \hline
Facebook                     & /data/com.facebook.katana/databases/threads\_db2          & 478             & 0                       \\ \hline
Facebook Messenger           & /data/com.facebook.orca/databases/threads\_db2            & 703             & 213                     \\ \hline
Kik                          & /data/data/kik.android/databases/.kikDatabase.db          & 431             & 1                       \\ \hline
Line                         & /data/jp.naver.line.android/databases/naver\_line/LINE.db & 211             & 205                     \\ \hline
Viber                        & /data/com.viber.voip/databases/viber\_messages/Viber.db   & 219             & 205                     \\ \hline
Telegram                     & /data/org.telegram.messenger/files/cache4.db/Telegram.db  & 217             & 204                     \\ \hline
\end{tabular}
\caption*{\label{tbl:apps-socialmedia-db}To capture data from social media databases an app requires the permissions READ\_EXTERNAL\_STORAGE, WRITE\_EXTERNAL\_STORAGE. 5635 (87.62\%) apps within the corpus request for both the permissions.}
\end{table*}

\paragraphb{Creation of fear and compliance} There are two ways apps in our corpus capture users' call data: (1) by extracting stored call history and logs; or (2) through actively recording user calls. We discovered that a substantial 4528 (53.77\%) apps of our corpus possess call logging capability. This allows adversaries to access detailed information, including call types, contacted phone numbers, associated contacts, and even call locations. To further the intrusive surveillance, about 28.88\% (2432) of stalkerware samples in our corpus perform call recording. The recording is triggered using the PhoneStateListener API, which detects changes in the phone state such as idle, incoming, or ongoing calls. Cross-referencing call data with contacts enables stalkerware to map the user's social network. This is supported by the fact that 1259 (14.95\%) app samples periodically query the Android device's contact database to capture various contact details, including ID, phone number, display name and, even email addresses. Out of 1259 apps that monitor contacts, 1094 apps target Android versions older than Android 9. The decline in contact monitoring could be due to the shift in storage of contacts from devices onto social media platforms or Google services.

\paragraphb{Omnipresent surveillance} Screen capturing capability can further aid abusers in maintaining constant surveillance over their victims' lives. Even if an app uses encryption, taking screenshots may effectively bypass it. 3636 (43.18\%) app samples from our corpus periodically capture screenshots using the MediaProjection API. To extend the surveillance further, even when the user is not actively using their device, a significant portion of apps in our corpus record ambient noise or capture photos using the device camera without the consent or knowledge of the victim. 2985 (35.44\%) apps in our corpus use the device camera to take a photo when triggered with specific events like if a remote command is received. And this is concerning given that 2172 (25.79\%) apps taking photos target Android versions 9 and above. A deeper dive revealed that numerous apps use the AudioSource attribute of MediaRecorder to capture more than phone calls. They record ambient sounds, suggesting they may catch nearby conversations even if the victim is not actively on a call. For example, if we look at the code in~\cref{lst:mediacapture} the AudioSource.MIC is used indicating that the app can capture any sound in the vicinity of the device. Also the use of AudioSource.VOICE\_RECOGNITION points out the alarming possibility that the app is trying to optimally pick up voice conversations; since this attribute is usually intended for voice recognition, which is used for processing and transcribing voice or interpreting voice commands. About 2160 (25.65\%) app samples from our corpus perform audio recording using AudioSource.VOICE\_RECOGNITION, out of which 1820 apps target Android versions 9 and above.

\begin{lstlisting}[language=Java, caption=MediaRecorder API, label=lst:mediacapture]
import android.media.MediaRecorder;
        MediaRecorder mediaRecorder = mRecorder;
        if (mediaRecorder == null) {
        double maxAmplitude = (double) mediaRecorder.getMaxAmplitude();
            android.media.MediaRecorder r3 = com.as.monitoringapp.VoiRe.LocalVoipRecord.mRecorder
            android.media.MediaRecorder r4 = com.as.monitoringapp.VoiRe.LocalVoipRecord.mRecorder
            android.media.MediaRecorder r2 = com.as.monitoringapp.VoiRe.LocalVoipRecord.mRecorder
            android.media.MediaRecorder r0 = new android.media.MediaRecorder     // Catch:{ Exception -> 0x01c9 }
                Log.e("CallingRecord_Type", "MediaRecorder.AudioSource.VOICE_CALL");
                Log.e("CallingRecord_Type", "MediaRecorder.AudioSource.MIC");
                Log.e("CallingRecord_Type", "MediaRecorder.AudioSource.VOICE_RECOGNITION");
                Log.e("CallingRecord_Type", "MediaRecorder.AudioSource.VOICE_COMMUNICATION");
\end{lstlisting}
\subsection{Evolution of Capabilities}
Figure \ref{fig:caps2} demonstrates the trends in stalkerware capabilities across Android apps targeting API levels between 28 and 33. From the figure, we can see that location capability is prevalent in stalkerware apps targeting all API levels. App samples that target API level 33 in our corpus consist of apps that focus primarily on location and phone call capabilities. This indicates that location tracking is an important feature for stalkerware apps. A high proportion of stalkerware apps targeting API level 28 in our dataset (85.3\%) perform SMS monitoring. This reflects the presence of family monitoring apps that use SMS commands as part of their functionality.

\begin{figure*}
\hspace*{-1.5cm}
\centering
    \includegraphics[width=0.98\linewidth]{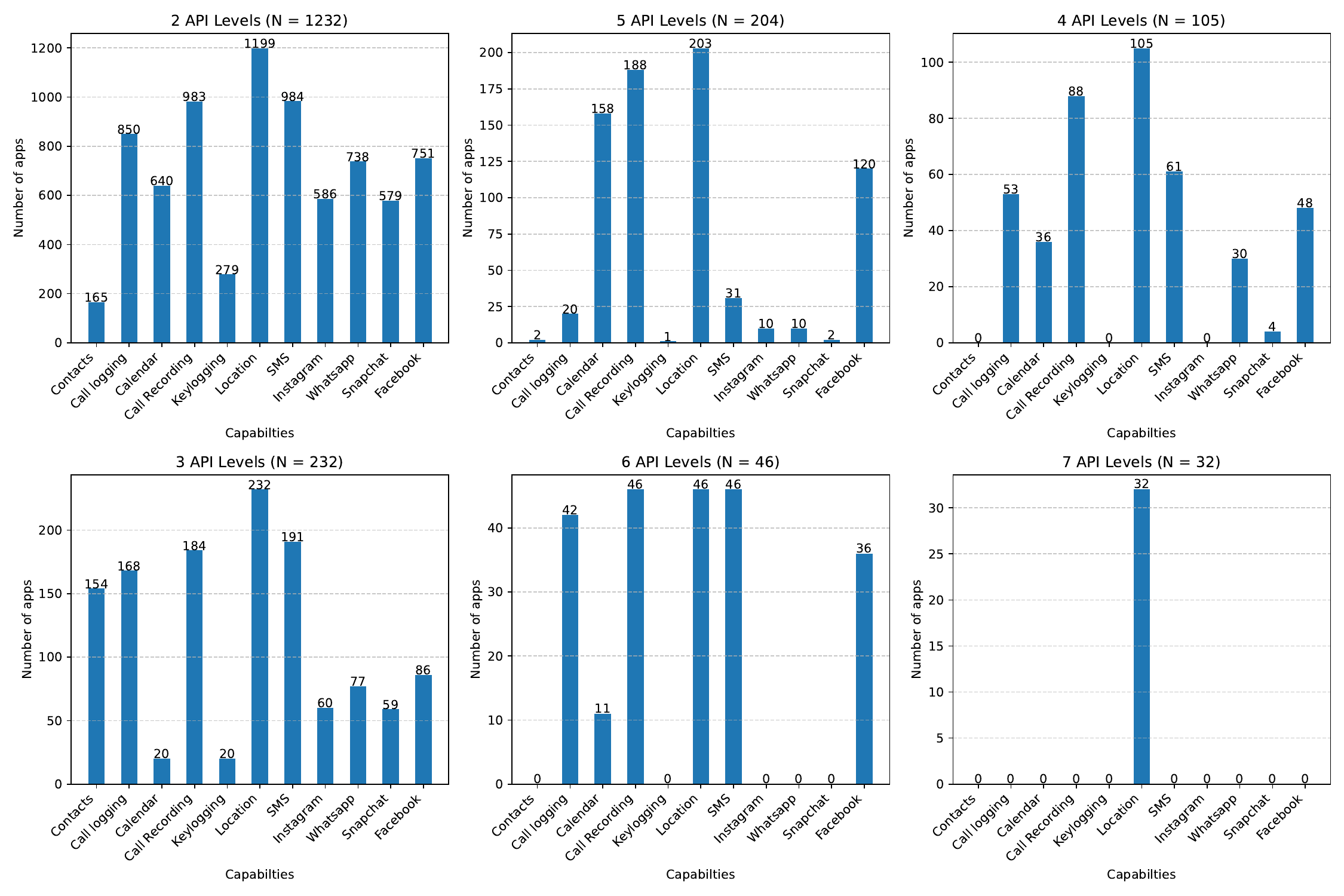}
    \caption{Capabilities possessed by apps that target different API levels across their different versions. This figure shows the raw count of all different versions of all such apps.}
    \label{fig:cap_count}
\end{figure*}

In our corpus, 87 unique apps have versions that target different API levels. For example, the app `AllTrackerFamily' has different versions of the app that target API levels 9, 27, 28 and 30. The most common capabilities possessed by all such apps are summarized in \cref{tbl:evolution}. Different versions of an app do not necessarily have the same capabilities. For example, \cref{fig:cap_count} highlights the capabilities possessed by the 46 apps that target 6 different API levels. All these 46 apps targeting 6 different API levels are different versions of the same unique app, yet we can see from the bar chart that the capabilities within those versions are different. The recent versions of this app (target API level 23 and above) do not monitor calendar events as opposed to the older version. Within our corpus, we also have examples of apps that possess the same capability across all different versions. For example, \cref{fig:cap_count} shows that 42 different versions of 2 unique apps that target 7 different API levels only perform location monitoring. 

\subsection{Corroborating Findings with Other Analyses}
\paragraphb{Taint Analysis with FlowDroid}
We used FlowDroid~\cite{arzt2014flowdroid} on 50 randomly sampled unique apps from our dataset, to trace taintflow for location, call logging, calendar, and contacts capabilities from our taxonomy. For the definition of sources and sinks, we used the SuSi tool~\cite{SuSi}. The apps from our random sample target SDK versions 28, 29, and 31. The results, as presented in~\cref{tbl:flowdroid}, indicate that our query-based method yields comparable outcomes to those obtained with FlowDroid. Notably, we observed instances where FlowDroid produced false negatives, failing to detect location and contact capabilities in certain apps. However, these capabilities were explicitly stated in user-visible strings such as installation agreements and warnings. A manual analysis of these apps further corroborated this observation.
A plausible explanation for these discrepancies could be attributed to the implicit calls commonly employed in stalkerware apps~\cite{DiscoveringFlawsinStaticAnalysisTools}. While not seeking to discredit tools like FlowDroid, our findings suggest that its analysis may benefit from enhancements to address such non-linear control flow, particularly prevalent in stalkerware applications.

\begin{table}[!th]
\centering
\caption{Capabilities detected using Taint Analysis} 
\label{tbl:flowdroid}
\resizebox{0.7\columnwidth}{!}{%
\begin{tabular}{cccc}
\toprule
Capability& FlowDroid  & Query-based approach \\ \midrule
Location  & 19   & 23  \\
Calendar  & 3   & 3 \\
Contacts  & 17  & 18 \\
Call logging  & 7   &  7   \\ \bottomrule
\end{tabular}%
}
\end{table}

\paragraphb{Dynamic Analysis}
We set up a controlled Linux-based Docker container to isolate the analysis environment from external influences. For testing, we used the Android Debug Bridge (ADB) to emulate different Android devices with Android versions 9 and above. The ADB provides detailed control and visibility over the emulator's operations, facilitating a comprehensive investigation. 
We then manually installed every unique APK sample from our data that targets Android 9 and above, on the emulator matching the targetSDKVersion of the APK sample, to imitate the real-world scenario where a user installs the app on a target device. We also observed the behavior of these apps, including the instructions provided during installation and the dashboards that list the capabilities visible to the adversary (we only observed the dashboards for apps that offer free trials or have no upfront subscriptions). 

\section{Android Platform Model Evolution}
\label{sec:changes}
In this section, we discuss the changes to the Android Platform Model that may impact stalkerware capabilities and what we observe in our corpus about them. We discuss the reactions of stalkerware apps to these changes in \cref{sec:resilience}.

\subsection{Android 6}
Android 6 (API level 23) introduced Doze and App Standby \footnote{https://developer.android.com/training/monitoring-device-state/doze-standby} to optimize power savings. Doze and App standby restrict background activities and network access of apps by interrupting the function of AlarmManager, JobScheduler and WorkManager APIs. Stalkerware apps in our corpus largely use AlarmManager and JobScheduler to capture user data in the background. In subsequent Android releases, more restrictions have been added to the App Standby buckets.

\subsection{Android 9}
\paragraphb{Call logging}
Android 9 introduced the \texttt{CALL\_LOG} permission group which basically contains all the permissions required to perform call logging.\footnote{\url{https://developer.android.com/about/versions/pie/android-9.0-changes-all}} 
This change was intended to give users better control and visibility to apps that need access to sensitive information about phone calls, such as reading phone call records and identifying phone numbers. Since these permissions are granted at runtime it gives an opportunity for users to deny the permission which would impede call logging capabilities of a stalkerware app. 

\paragraphb{Device administrator apps}

With the introduction of Android 9 for device administrator apps, policies related to control of screen lock, camera, and device password were deprecated. In the following Android releases, even more policies like forced data wipeout or device lock were also deprecated, and device administrator was officially considered as deprecated. However, apps targeting Android 9 and above (up to the most recent Android release) still support enabling apps to be device administrators and making them exempt from battery optimization and other restrictions. Currently, 2674 (31.75\%) apps in our corpus ask to be enabled as device administrator apps mainly to evade restrictions related to battery optimization. The functionality of these apps would be severely impacted by the deprecation of the device administrator.

\subsection{Android 10}
\paragraphb{Scoped storage}

To give users more control over their files and limit file clutter, apps that target Android 10 (API level 29) and higher are given scoped access to external storage, or scoped storage, by default.\footnote{\url{https://developer.android.com/training/data-storage\#scoped-storage}} This change is likely to impact the apps that capture data from social media apps. Since if a user has chosen a specific external storage directory to store files and media from a social media app, due to the scoped storage stalkerware apps will not be able to access those files.

\paragraphb{Location (permission) only while in the app}
To provide more control over location data Android 10 introduced a new permission option -- users can now allow an app to access location only while the app is actually in use (running in the foreground).\footnote{\url{https://developer.android.com/about/versions/10/highlights}} By choosing this option a user can stop the frequent location access of an app which is the most important capability for apps in our corpus with 90.66\% apps performing location tracking.

\subsection{Android 11}
\paragraphb{Temporary permissions}
Android 11 introduced an option for users to grant only one time permissions to and app using ``Only this time'' option. Android 11 also introduced auto-reset for runtime permissions used to access Location, Camera, Accounts or make Phone Calls, perform call logging, and more. So, if the user did not interact with the app with permissions for a few months, all the runtime permissions would be completely reset. Both these changes would require users to regrant permissions which could impact the multiple capabilities of stalkerware apps especially if during the installation all these permissions were covertly granted by the adversary.

\paragraphb{Phone number permission} Android 11 replaced the permission \texttt{READ\_PHONE\_STATE} that had a broader scope with the permissions \texttt{READ\_PHONE\_NUMBERS}. This change helped narrow the scope of accessing sensitive information, such as phone numbers. All the apps in our corpus that perform call logging and targeting Android 11 and higher were impacted by this change in permission.

\subsection{Android 12}
\paragraphb{Microphone and camera indicators}
On devices running Android 12 or higher, when an app accesses the microphone or camera, an icon appears in the status bar. This feature ensures that the user gets a visual indicator of being recorded. Within our corpus, 5510 and 6075 apps request camera and audio recording permissions, respectively. These apps will now have a microphone or camera icon in the status bar when recording.

\paragraphb{App hibernation}
 If an app targets Android 12 and the user hasn't interacted with the app for a few months, the system auto-resets any granted permissions and places the app in a hibernation state. This change would impact all the apps in our corpus, since in most cases the adversary gets specific instruction on how to grant all the necessary permissions that the app needs to function properly.

\section{Resilience Tactics of Stalkerware} \label{sec:resilience}
So far, we discussed the changes to the Android platform model and how they may affect stalkerware. In this section, we discuss the tactics used by stalkerware to avoid being affected by restrictions on capabilities resulting from changes to Android.

\subsection{Installer/Side Loading apps}  \label{sec:resilience:installer}
Our corpus contains 765 (9.08\%) apps that do not request any permissions from the user. These apps are not flagged by  Google Play Protect, as no dangerous permissions are requested. After dynamically analyzing (these) apps, we found that these apps act as installation guides or tutorials for the actual stalkerware apps. These apps provide detailed instructions on how to disable settings and notifications to keep stalkerware undetected for the specific Android version, how to install the actual stalkerware app, and grant the required permissions.

\subsection{Rooting and Device Administrator apps} \label{sec:resilience:rooting}
As in a rooted device, stalkerware can directly gain access to encrypted databases, using shell commands, commonly used for social media apps. Therefore, it is not surprising that 3716 (44.12\%) apps ask the installer to make sure that the device is rooted for the stalkerware apps to function correctly. A large portion of these apps advise the adversary (who performs the installation of the app) to search the web for the procedure to root their device based on the device manufacturer. Apps that are enabled as device administrator apps are exempt from battery optimization restrictions specifically imposed by Doze and App Standby in Android. We found that 2674 (31.75\%) apps in our corpus provide instructions to be enabled as device administrator apps during the time of installation. This also protects them from being uninstalled.

\subsection{Instructions to the adversary} \label{sec:resilience:instructions}
To ensure that the stalkerware apps function correctly, the adversary receives a series of instructions during the installation of the apps on the victim's device. The following subsection details these instructions.

\paragraphb{Disabling Battery optimization} Since battery optimization affects the majority of apps in our corpus, this is the most observed and significant instruction that we noticed. We found a few different variations of the instructions related to battery optimization. For example, a more generic and commonly observed instruction looked like \textit{``In order to monitor your location during low battery the app needs to be allowed to run in the background. 
On the next screen select `All Apps' then select App's name and change to Don't optimize."} For a few apps, we also observed very detailed instructions on how to add the installed stalkerware to the system apps list or protected apps for several different device manufacturers. This is probably to ensure that they remain active and function properly, especially while running in the background. For example,\textit{``On Huawei phones please add the App to the Protected apps list. Go to Settings -> Advanced settings -> Battery manager -> Protected apps"}.

\paragraphb{In response to specific changes in the Android platform model} To make sure that the app functions correctly, stalkerware apps provide instructions on how to handle specific restrictions imposed on them due to updates in the Android operating system. For example, we found the following instructions in our corpus: \textit{``Starting with Android 10, system security has been improved, which has impacted the “Screen streaming” feature. Screen streaming permission can no longer be granted permanently – it should be granted in settings after each restart of the target device."} or \textit{``Due to the features of the Android system, to create screenshots, it is necessary to display the application icon in the upper curtain. If this icon bothers you, you can hide it as follows: Settings -> Applications and notifications (Apps) -> NS Cloud -> Notifications -> Turn off notifications."}

\paragraphb{SMS commands} As discussed in Section \ref{sec:commands}, a significant amount of stalkerware apps use SMS commands to start capturing specific user data. For example, we see apps providing a list of commands to execute different capabilities with disclaimers such as \textit{ ``This app functions as a receiver for SMS commands. So you can control this phone by sending simple control messages. (e.g. start audio recording)."}

\paragraphb{Warnings about other apps} A small portion of the apps in the corpus provide warnings if the victim's device already has another tracking app, antivirus app, or task-killing app installed. Some stalkerware apps go as far as maintaining an exhaustive list of popular antivirus, task killer, and tracking apps to flag them and instruct the adversary to uninstall them during the installation phase. One such example from our corpus is: \textit{``We have detected an antivirus application installed on the phone. We recommend that you uninstall this application, or you can add this app to the whitelist.''}

\paragraphb{Other Instructions} One of the most common instructions we encountered in our corpus was how to enable accessibility services. We also observed warnings about the installation of Google Play services (if absent) as some of these apps use APIs provided by Google Play services. For example, API \texttt{com.google.android.gms.ads} for Mobile Ads or \texttt{com.google.android.gms.location} for location capture. We also observed that most of the apps in our corpus have detailed instructions on how to disable Play Protect \footnote{This is because Google Play Protect issues warnings if an app seems harmful and tries to capture a lot of personal data.}. Some apps from the corpus not only have detailed instructions on the app's user interface but also provide links to public videos that explain every installation step with visuals. 

\subsection{Covering the tracks:}
\paragraphb{Data Deletion}
We discovered that some apps in our corpus hide their activity by deleting all related files from the device's storage. They may use a specific naming convention to identify these files, such as a keyword or a timestamp. The apps can then delete all files that match this convention periodically or after certain conditions are met as shown in \cref{lst:deletion}. The deletion could be to remove an older version of the app or temporary files. However, within the apps we examined, deletion occurs after the execution of background services that capture user data, like location coordinates, suggesting it is used to erase evidence of data collection.

\label{lst:deletion}
\begin{lstlisting}[language=Java, caption=Deletion of files associated with the app, label=lst:deletion]
 public void run() {
            try {
                File downloadDirectory = new File(String.valueOf(Environment.getExternalStorageDirectory().getAbsolutePath()) + "/download");
                String[] fileList = downloadDirectory.list();
                int length = fileList.length;
                for (int i = 0; i < length; i += p128d18b9c3.SYSTEM_IMEI_INDEX) {
                    String fileName = fileList[i];
                    if (fileName.indexOf("mobistealth") > -1) {
                        new File(downloadDirectory + "/" + fileName).delete();
                    }
                }
            } catch (Exception e) {
            }
        }
 \end{lstlisting}
\paragraphb{Anti-reverse engineering strategies}
We encountered the usage of several anti-reverse engineering strategies like code obfuscation, dummy functions, redundant variables, and even dead code employed throughout our corpus.

\subsection{Communication with the backend}\label{sec:commands}
Stalkerware apps typically need to communicate with a backend to exfiltrate data and provide monitoring capabilities to the attacker. Performing capture on command also prevents the apps from capturing data too frequently and thus being flagged as battery draining. Therefore, we investigate how stalkerware apps communicate with their backend to receive instructions about performing certain capabilities. 

We found that a large number of stalkerware apps receive commands via SMS and the network, allowing adversaries to trigger specific capabilities at specified times and giving them greater control over the targeted device. These commands serve as guides, directing the application in strategies for capturing data. These commands contain details such as what specific data to capture and the frequency or time next capture or exfiltration of data. Within our corpus, 3268 (38.80\%) apps receive SMS commands to perform a relevant action. We also found that all apps that access network libraries make GET requests that can be used to fetch commands from the server, as shown in~\cref{lst:networkcommand}. 

\begin{lstlisting}[language=Java, caption=Retrieving Commands , label=lst:networkcommand]
private BufferedReader sendHttpGetRequest(String urlString) {
        try {
            System.out.println("[StealthCommandReceiver]: " + urlString);
            return new BufferedReader(new InputStreamReader(new URL(urlString).openConnection().getInputStream()));
        } catch (Exception e) {
            return null;
        }
    }

    private BufferedReader checkServerForCommands(String imei) {
        BufferedReader buffReader = null;
        String url = String.valueOf("http://mobistealth.com/cmd_issuance.php?imei=") + imei;
        for (int retryCount = 0; retryCount < 3 && (buffReader = sendHttpGetRequest(url)) == null; retryCount++) {
            try {
                Thread.sleep(1000);
            } catch (InterruptedException e) {
            }
        }
        return buffReader;
    }
\end{lstlisting}
The above code includes methods such as sendHttpGetRequest and checkServerForCommands, which facilitate connecting to a remote server and fetching commands from the specified remote server of mobistealth, based on a device's IMEI number. 

While in most cases, the instructions on how to issue commands to execute a particular capability are detailed on the stalkerware app's website or the adversary dashboard, there are a few apps that also have instructions for adversaries within the app installed on the victim's device. At least, 2599 (30.86\%) apps in our corpus have these instructions present on the app on the victim's device. We consider this number as a lower bound since we have only considered instructions provided to the adversary that are in English when counting. We observed that 1448 (17.19\%) apps within our corpus primarily use a language other than English throughout their user interface elements. This suggests that these apps may be designed specifically for a particular user demographic. We encountered the use of 103 distinct languages in the user interface elements throughout the apps within the corpus.

\section{Discussion}\label{sec:discussion}
\begin{table*}[th]
\caption{Frequency of APIs and Permissions for each capability.}
\small
\label{tbl: capabilities}
\begin{tabular}{ccccccc}
\toprule
Capability                                                                        & Method of Access          & Samples (\%) & Unique & Permissions &  Samples [Unique]\\ \midrule
Keylogging                                                      & AccessibilityService API  & 600 (7.12\%)       &  72   &  BIND\_ACCESSIBILITY\_SERVICE   & 600 [72] \\ \midrule
Calender                                                                          & CalendarContract API      & 3523 (41.83\%)       &  836  & READ\_CALENDAR, WRITE\_CALENDAR   & 4695 [1440]  \\ \midrule
Contacts                                                                          & ContactsContract API      & 1259 (14.95\%)       & 474    & READ\_CONTACTS, WRITE\_CONTACTS                                                                                                         & 4717 [534]  \\ \midrule
Ambient Noise Recording                                                                          & MediaRecorder API       & 2160 (25.65\%)       & 312   & RECORD\_AUDIO  & 6075 [2155]        \\ \midrule
\multirow{4}{*}{Location}                                                         & LocationManager API       & 7177 (85.23\%)    & 2729   & ACCESS\_FINE\_LOCATION,                                                                                        & 7177 [2729]   \\ 
                                                                                  & FusedLocationProvider API & 1250 (14.84\%)        & 436       &  ACCESS\_COARSE\_LOCATION    \\ 
                                                                                  & Geofencing API            & 1014 (12.04\%)      & 393 &  ACCESS\_BACKGROUND\_LOCATION & 3197 [695]
                                                                                  \\
                                                                                  & TelephonyManager API      & 2707 (32.14\%)        & 1818      \\ \midrule
\multirow{4}{*}{Phone calls}                                                      & CallLog API               & 4528 (53.77\%)       & 1554  & READ\_CALL\_LOG, WRITE\_CALL\_LOG,                                                                                                            & 7094 [2671]           \\ 
                                                                                  & PhoneStateListener API    & 2432 (28.88\%)       & 1310 & READ\_PHONE\_STATE,      \\ 
                                                                      &&&&READ\_PHONE\_NUMBERS     \\       & MediaRecorder API         & 6026 (71.55\%)       & 2212 &RECORD\_AUDIO,READ\_PHONE\_STATE   & 6075 [2155]         \\ \midrule

\multirow{3}{*}{Messages}                                                         & SMS Provider              & 5932 (70.44\%)       & 2924   & READ\_SMS, RECEIVE\_SMS                                                                                                                 & 5935 [2212]  \\ 
                                                                                  & SmsMessage                & 4505 (53.49\%)       & 1398     \\ 
                                                                                  & MMS Provider             & 2955 (35.09\%)      & 1518 & RECEIVE\_MMS     &  2955 [1518]   \\ \midrule

\multirow{6}{*}{Data Exfiltration}                                                & HttpURLConnection API     & 7752 (92.05\%) & 1993  & INTERNET                                                                                        & 7976 [3085] \\ 
                                                                                  & OkHttpClient API          & 3049 (36.21\%) & 314    \\ 
                                                                                  & Retrofit API              & 208 (26.22\%) & 149      \\ 
    & SmsManager API          & 2760 (32.77\%) & 418 & SEND\_SMS    & 3276 [418] \\
    
                                          \bottomrule
\end{tabular}
\label{tbl:APIcap}
 \caption*{Please note that the percentage values for `Samples' column are with respect to the total corpus of size 8421.}
\end{table*}

In this section, we reflect on our findings and discuss them in the broader context of the stalkerware ecosystem.

\subsection{Have Changes Thwarted Stalkerware?}
Our findings do not support the hypothesis that changes in the Android platform model have had a substantial effect on reducing stalkerware capabilities. As discussed in the previous section, the specific changes themselves caused observable reactions in the stalkerware app, as inferred from the instructions to stalkers found in installer apps (\cref{sec:resilience}). An example of this is Google Play Protect warning about harmful apps and the resulting installer apps detailed instructions for stalkers about how to disable Play Protect. However, these changes to Android do not appear to have significantly reduced the capabilities offered by recent stalkerware. \cref{fig:caps2} shows that while there are differences in capabilities coverage across API levels --- due to idiosyncrasies of apps in those levels and the non-uniform distribution across levels --- none of the capabilities appear to have been removed in recent versions. If a change at some API level had effectively prevented stalkerware from having a specific capability, we would expect stalkerware apps targeting the latter version not to possess such a capability. %

Even when platform-level changes are introduced to curb malicious behaviors, their enforcement can be delayed in practice. For instance, stalkerware developers may circumvent restrictions by setting the app’s targetSdkVersion to an older API level, effectively opting out of new limitations. However, Android has recently moved toward stronger enforcement: as of Android 15, newly published and updated apps are required to target more recent API levels.\footnote{\url{https://developer.android.com/about/versions/15/behavior-changes-all\#minimum-target-api-level}}

This change, however, primarily applies to apps distributed through the Google Play Store. Since October 2020, Play Store policy updates have explicitly banned the distribution of stalkerware, forcing developers to rely on alternative distribution channels, such as third-party app stores, direct downloads, or sideloading, mechanisms that are not always subject to the same enforcement mechanisms.

Taken together, while platform policy and SDK targeting enforcement represent important steps forward, their real-world impact may remain limited in the short term.

\subsection{Why Have Privacy Updates Not Solved It?} %
Consider the list of some important changes introduced by Android with the launch of newer Android versions:
\begin{enumerate}[leftmargin=1.75em]
    \item Since Android 9, nearly every new Android OS update has introduced features to give users more control over location services, for example, new permission options, one-time permissions, and a privacy dashboard. But during installation, apps can still be granted \textit{``Allow all the time''} permission. In the case of stalkerware, an adversary installing the stalkerware on the victim's phone can easily grant this permission, and most stalkerware apps in our corpus instruct the person installing the app to do so. This also applies to other permission changes mentioned in~\cref{sec:changes}.
    \item Android 6 introduced Doze to reduce battery consumption. The apps in Doze cannot access the network or use the JobScheduler or AlarmManager APIs. However, apps that are on the Doze exemption list do not have to face these restrictions. Users can manually configure the list of exempt apps from \texttt{(Settings > Battery > Battery Optimization)}, and we found that 853 apps in our corpus provide these instructions during installation. As per Android documentation, apps that are on the Doze exempt list are also exempt from the restriction of App Standby Buckets.\footnote{\url{https://developer.android.com/topic/performance/appstandby}} 
    \item Android 9 introduced App Standby Buckets to impose restrictions on apps that do not interact with the user or execute tasks in the background. According to the latest Android documentation, the App Standby Bucket restrictions can be evaded if apps qualify for exemptions such as if they possess permissions like \texttt{ACCESS\_BACKGROUND\_LOCATION} or \texttt{USE\_EXACT\_ALARM} which can be granted to the app during installation.
    \item Since Android 10 device administrator apps are considered deprecated. However, users can still enable apps (as of Android 14) to be device administrators by directly using device settings. 
    \item The Android accessibility service is intended to enhance the user interface to help users with disabilities or who may temporarily be unable to fully interact with a device. However, the apps in our corpus use this functionality to capture screen content as well as toasts and notifications from social media apps. Despite the potential for misuse, Android cannot restrict accessibility services because doing so will impact users who rely on these features for accessibility. It is also complex to distinguish malicious use of accessibility services from legitimate use. For instance, activities like screen reading, accessing the content of app notifications, and user input monitoring have legitimate uses for assisting individuals with disabilities.
   
\end{enumerate}

This suggests two broad reasons why privacy updates have not thwarted stalkerware.

\paragraphb{Maintaining compatibility and accessibility}
The accessibility services for Android examplifies a fundamental trade-off inherent to any changes to the Android platform model. To maintain accessibility services for users who need them, Android cannot restrict their access. Therefore stalkerware apps can use them for functionality. Although numerous changes are introduced to harden the permission system, which facilitates the elimination of features that compromise user privacy, these changes are usually not enforced right away (or at all) to maintain backward compatibility. In this sense, there is little Android can do to thwart stalkerware if it must maintain backward compatibility and essential functionality like accessibility services. 

\paragraphb{The stalkerware adversary}
Another reason stalkerware can circumvent the hardening of the Android platform model is its unique adversary. The stalkerware is unlike more traditional adversaries for mobile devices, such as a remote adversary that tries to get access to sensitive information on the device through traditional malware. This is because the stalkerware adversary in many instances may have physical access to the victim's device at various times~\cite{chatterjee2018spyware,woodlock2017abuse}.  This enables stalkers not only to install stalkerware apps surreptitiously but also to give stalkerware apps more access than regular apps. For example, stalkers may root the device to facilitate stalkerware capabilities, or they may install the app as a device administrator app. They may ignore installation-time warnings and disable critical protection such as Google Play Protect or other anti-malware software that may be present on the device. Furthermore, even if the victim is aware that stalkerware is present on the device, they may be unable or hesitant to remove it.

\subsection{Other Avenues for Mitigation}\label{sec:mitigation}
\paragraphb{Protecting sensitive information on rooted devices}
Recall that a significant portion of stalkerware apps access databases from social media and messaging apps such as WhatsApp, Instagram, and Facebook. These databases are encrypted, but if the user's device is rooted, stalkerware apps can bypass encryption. Since a significant portion of the apps in our corpus access these databases \cref{tbl:APIcap}, this suggests that stalkerware apps anticipate running on a rooted device. Therefore, straightforward use of encryption is not sufficient as a defense because the decryption key must be available for use by legitimate apps.

\paragraphb{Granularity of location tracking}
More than 90\% of the apps in our corpus track the location of the user. In Android 8 (API level 26), modifications were made to location retrieval for background apps, affecting commonly used tools such as the Fused Location Provider, Geofencing, and LocationManager API—frequently employed by stalkerware apps. These changes restricted location updates to only a few times per hour. This adjustment has proven effective in mitigating the potential misuse of stalkerware apps by curbing unauthorized real-time location tracking of victims.\footnote{\url{https://developer.android.com/about/versions/oreo/background-location-limits}} However, the batched version of the impacted APIs will allow the app to receive the user's location more frequently than the non-batched API. Updates in batches are received only a few times per hour. Android 11 (API level 30) introduced more user control of background location tracking by removing the option to grant location permission \textit{``all the time''} from the system permissions dialog. Although uninterrupted background location permission can still be granted to an app easily through the settings page.
A related feature idea that could offer protection for location tracking in the context of stalkerware would be to provide users with the ability to reduce the granularity of updates on a per-app basis so that users could elect to provide cached or stale location information (to stalkerware apps).

\paragraphb{Delay in permission granting}
All the apps in our corpus request a variety of permissions that are considered dangerous as per Android documentation. In the case of stalkerware, these permissions are granted to all during the installation time, and in most cases the apps provide instructions to grant these permissions with the option \textit{ ``Allow all the time''} present in the Android device settings. The delay in granting such permissions if the option \textit{ ``Allow all the time''} is chosen might prevent the stalkerware apps from receiving permissions to keep functioning covertly to capture user data.

\section{Study Limitations \& Ethics}\label{sec:limits}
\paragraphb{Limitations}
A limitation of any large-corpus study (including ours) is the inability to establish with certainty whether code mapping onto a specific capability of behavior is executed at runtime. Along with this comes the fact that our analysis only provides a lower bound on the number of samples in our corpus that engage in a specific implementation of a capability, as there could (in theory) be other ways to implement it that our methodology did not uncover. To mitigate this, we perform dynamic analyses, taint analyses, and manually inspect numerous app samples. Through this, we are able to confirm many of our findings. Nevertheless, obfuscation and anti-reverse engineering measures impede analysis. Reinforcing this point, we found that all the apps in our corpus use meaningless variable and method names, as well as redundant and dead code, to complicate the analysis process and evade detection by security tools. However, this kind of obfuscation does not lead to over-(or under-)counting the number of samples that possess a specific capability. 

\paragraphb{Ethical considerations}
Our primary motivation in studying stalkerware capabilities and behaviors is to aid in the design of defenses to thwart stalkerware and thus reduce harm. However, we are aware of potential indirect harms from studying stalkerware and have taken steps to mitigate this. For example, elucidating details about the fine-grained capabilities of stalkerware may aid abusers in their search for stalkerware apps. Overall, we believe that the benefits of improving the understanding of stalkerware, such as uncovering new insights to design effective defenses, outweigh the potential harms that stem from a study of stalkerware such as ours.

\section{Conclusions \& Future Directions}\label{sec:conclusions}

We perform a comprehensive large-scale quantitative study of Android stalkerware apps' capabilities with the goal to analyze whether the evolution of Android platform model has impeded stalkerware. We combine multiple analysis techniques to systematically quantify the stalkerware capabilities found in our corpus and correlate them with the fine-grained changes introduced in recent Android versions.

While some Android platform changes affect stalkerware apps and result in observable behavior changes in subsequent versions of some apps, the stalkerware app ecosystem seems to have largely adapted without much (if any) loss of functionality. 

The problem, arguably, is that most of the hardening of the Android platform (not all of which is intended to thwart stalkerware) can be circumvented because stalkerware adversaries potentially have physical access to the device and former versions of updated APIs continue to operate to allow backward compatibility. 
 
We hope that our findings inform adversarial models of stalkerware and aid in developing effective defenses. 

Future investigations should seek to better understand how information flows within the stalkerware ecosystem and the potentially differing behavior between stalkerware and dual-use apps.


\appendix
\end{document}